# Development and evaluation of a 3D annotation software for interactive COVID-19 lesion segmentation in chest CT

Simone Bendazzoli[1*], Irene Brusini[1,2*], Mehdi Astaraki[1,3], Mats Persson[4], Jimmy Yu[5], Bryan Connolly[5], Sven Nyrén[5,6], Fredrik Strand[3,5], Örjan Smedby[1], Chunliang Wang[1*]

[1] Department of Biomedical Engineering and Health Systems, KTH Royal Institute of Technology, Hälsovägen 11C, SE-14157 Huddinge, Sweden
[2] Department of Neurobiology, Care Sciences and Society, Karolinska Institute, 141 83 Huddinge, Sweden
[3] Department of Oncology-Pathology, Karolinska University Hospital, Solna, SE-17176 Stockholm, Sweden
[4] Department of Physics, KTH Royal Institute of Technology, Roslagstullsbacken 21, SE-10691, Sweden
[5] Department of Imaging and Physiology, Karolinska University Hospital, Solna, SE-17176 Stockholm, Sweden
[6] Department of Molecular Medicine and Surgery (MMK), Karolinska Institute, SE-17176 Stockholm, Sweden

## Abstract

Segmentation of COVID-19 lesions from chest CT scans is of great importance for better diagnosing the disease and investigating its extent. However, manual segmentation can be very time consuming and subjective, given the lesions' large variation in shape, size and position. On the other hand, we still lack large manually segmented datasets that could be used for training machine learning-based models for fully automatic segmentation. In this work, we propose a new interactive and user-friendly tool for COVID-19 lesion segmentation, which works by alternating automatic steps (based on level-set segmentation and statistical shape modeling) with manual correction steps. The present software was tested by two different expertise groups: one group of three radiologists and one of three users with an engineering background. Promising segmentation results were obtained by both groups, which achieved satisfactory agreement both between- and within-group. Moreover, our interactive tool was shown to significantly speed up the lesion segmentation process, when compared to fully manual segmentation. Finally, we investigated inter-observer variability and how it is strongly influenced by several subjective factors, showing the importance for AI researchers and clinical doctors to be aware of the uncertainty in lesion segmentation results.

## 1. Introduction

Since the end of 2019, a new coronavirus disease (COVID-19) has started spreading across the world, leading to a pandemic that still represents a global health crisis. Up to January 10th, 2021, almost 90 million COVID-19 cases have been reported across 223 countries and territories, and 1 926 625 of these cases represent confirmed deaths [1].

The most common clinical manifestations of COVID-19 consist of fever, cough and dyspnea [2-4]. Such respiratory problems are often reflected in the outcome of chest X-ray and thoracic CT exams, which reveal pneumonia either bilaterally in both lungs or—less commonly—unilaterally. Pneumonia is most often identified as a ground-glass opacity in the X-ray or CT scan. However, other types of features can also be identified from the images, such as mixed ground-glass opacities and mixed consolidation lesions, reticulations, and crazy-paving patterns, especially in the later stages of the disease [5-10]. These image features are more evident in CT scans, which, compared to X-rays, have the advantage of providing a three-dimensional—and thus more informative—view of the lungs. Therefore, the use of CT for detecting the presence, type and extent of COVID-19 lesions is particularly promising, not only for diagnostic purposes, but also for better understanding the disease progression and possible treatment responses [11]. However, COVID-19 lesions are often diffuse and can vary in shape, size and position [12]. Thus, the manual delineation of such lesions from CT scans can be an extremely challenging and time-consuming task, as well as strongly affected by inter-rater variability. These issues are particularly problematic when studying a disease that is as new and rapidly spreading as COVID-19, since the international research effort to deal with the pandemic is moving at an extremely fast pace.

---

[*] Contributed equally

In recent years, deep learning-based methods have shown very good results within the field of medical image segmentation. Therefore, a few studies published this year have also employed such architectures to perform COVID-19 lesion segmentation automatically [6, 12-18], which would overcome the issues of time inefficiency and subjectivity that come with manual segmentation. Most deep learning-based segmentation methods are usually trained in a supervised fashion, i.e. they require a sufficiently large training dataset that already includes ground-truth segmentation masks. However, as previously discussed, the availability of manually delineated ground-truth is still largely lacking for COVID-19 lesions. Thus, some deep learning-based studies have tried to tackle this issue by proposing computer-assisted alternatives to the standard manual segmentation process.

In the work by Shan et al. [12], a combination of a V-Net [19] with a bottleneck structure [20] was employed to perform automatic lesion segmentation. The problem of ground-truth generation was here approached by adopting a human-in-the-loop strategy: only a limited batch of training data is delineated fully manually, while the rest of the ground-truth segmentations is created by manually correcting the outputs provided by the proposed segmentation network. These iterative manual corrections were shown to lead to a better segmentation accuracy, as well as a significantly reduced annotation time compared to a fully manual approach. On the other hand, in the study by Fan et al. [6] a semi-supervised learning strategy was used for training their proposed *Inf-Net*. This approach is similar to the one used by Shan et al. [12], but it avoids the manual correction steps of the network's output: the training set is iteratively enlarged by training the architecture with bigger and bigger batches that are generated by the segmentation network itself. Another method for alleviating the burden of ground-truth creation was recently proposed by Yao et al.: *NormNet* [17], a pixel-level anomaly model that synthesizes normal CT images from abnormal ones by removing COVID-19 lesions. This model is trained to learn a decision boundary between normal healthy tissue and lesions added synthetically in the training images. The performance of such model was comparable to other supervised models. Finally, weakly supervised methods constitute another alternative to deal with unavailability of large scale labeled dataset. By embedding generative adversarial training into a segmentation network, *GASNet* [18] was developed to strengthen the performance of the segmentation network with only very small labeled data. While the generator synthesizes healthy images from unhealthy ones, the discriminator is trained to distinguish them, and the adversarial signals are integrated into a segmentation network.

The above-mentioned studies showed promising results, while also proposing interesting ways to integrate the radiologists' knowledge with automated AI systems. However, machine learning-based approaches have two main drawbacks: (1) a preliminary fully manual segmentation of a relatively large number of images is necessary for starting to train the networks; (2) programming skills are required in order to properly re-train and implement these models.

In our work, we developed a novel interactive tool for COVID-19 lesion segmentation, implemented within the software *Mialab* (http://mialab.org). The proposed tool aims at segmenting first the lungs and later the lesions from input CT scans, by alternating automatic steps (based on the use of level-set segmentation, as well as the integration of statistical shape models) with manual correction steps. This semi-automatic pipeline does not require any previous training using manually segmented CT scans. Moreover, the creation of a simple user interface facilitates the use of our tools for any user.

Our proposed pipeline aims at speeding up and facilitating the creation of ground-truth lesion segmentation masks by providing a user-friendly tool that does not require any additional programming effort. On the other hand, the ground-truth segmentations created with our method could also be employed as training datasets for later training more complex deep learning models in a supervised fashion.

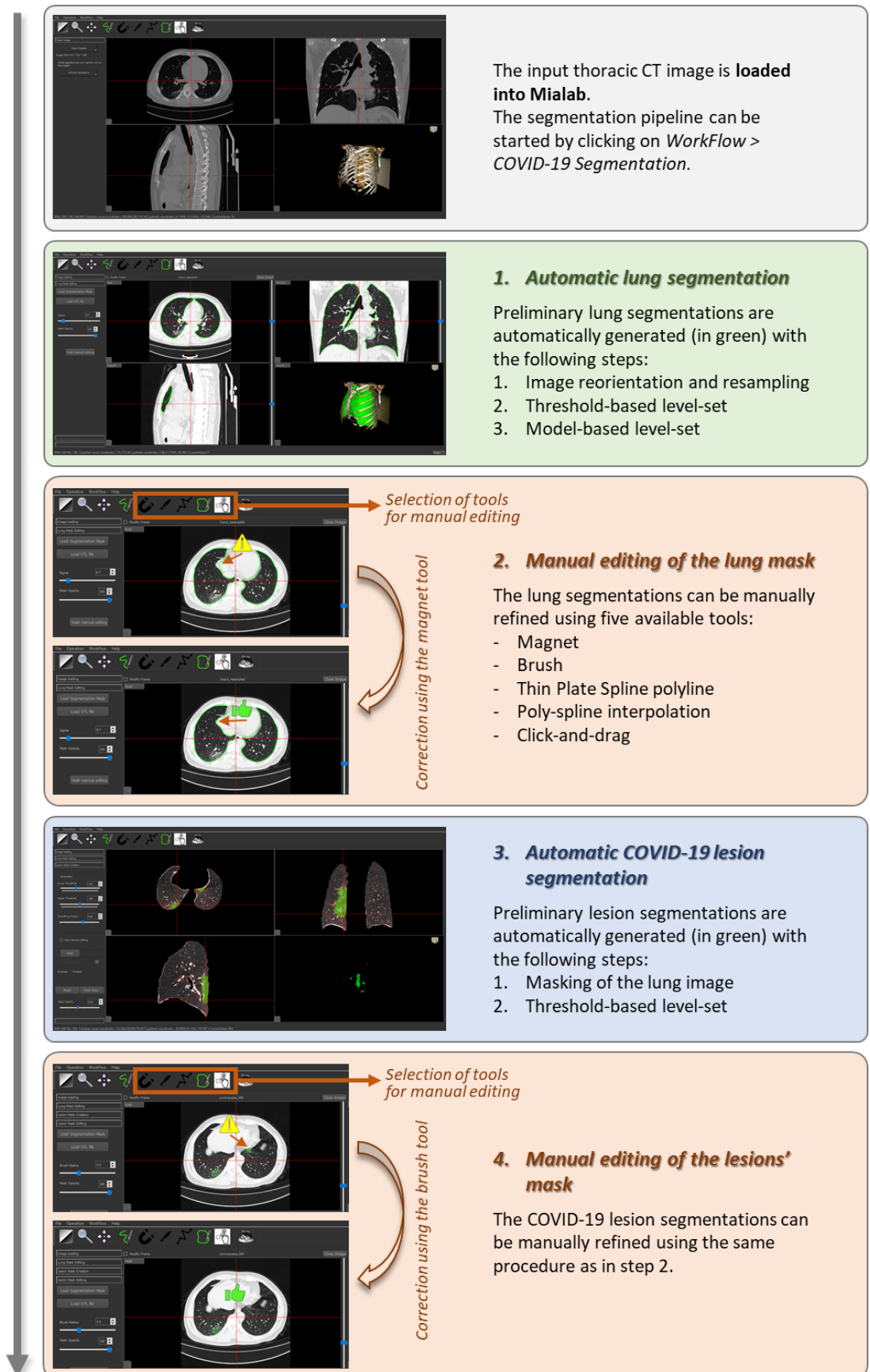

**Figure 1.** Schematic summary of the steps of the proposed pipeline for COVID-19 lesion segmentation. For each step, the graphic interface of Mialab is shown using an example thoracic CT scan as input to be segmented.

## 2. Materials and methods

The semi-automatic pipeline presented in this paper (see Figure 1) aims at identifying and segmenting COVID-19 lesions from chest CT scans. The workflow combines automatic segmentation methods with interactive editing and refinement, and it can be divided into four different sequential sections. The first one takes the chest CT scan as input and automatically produces a lung segmentation mask. The second section allows the user to refine manually the lung mask, and the third one takes the refined lung mask and performs an automatic segmentation to produce the COVID-19 lesion mask as output. Finally, the fourth section allows the user to perform a final editing of the COVID-19 lesions mask.

### 2.1 Automatic lung segmentation

The first section of the pipeline aims at producing a lung segmentation (volumes including both normal and abnormal lung parenchyma) and it can itself be divided into five main steps presented below: pre-processing, lung field estimation, initial model registration, right and left lung segmentation based on shape model fitting.

*2.1.1 Pre-processing*

Initially, a sequence of pre-processing operations is performed in order to transform the input CT image into a standard pre-defined image orientation and voxel resolution.

First, the image orientation is fixed according to the RAI matrix orientation (i.e. $X$ axis: Right to Left; $Y$ axis: Anterior to Posterior; $Z$ axis: Inferior to Superior). Then, the image is down-sampled to a 3x3x3 mm voxel spacing, using a triangle convolution. Lastly, a 1x1x1 isotropic resampling is performed, using linear interpolation.

*2.1.2 Lung field estimation*

The down-sampled image is then automatically segmented into the two lungs. A threshold-based level set segmentation method [21] is used for this purpose. The present method requires some parameters to be specified in order to automatically run the segmentation:
- A set of threshold values, which specify the voxel regions in which the segmentation is propagated. In this work, the lung regions to be segmented are identified in the range [–860, –200] HU.
- A curvature factor, which constitutes an internal constraint that determines the degree of smoothness and regularization in the level-set curve. Such curvature factor was set to 0.6.

The produced output is a lung distance map, defining the lung region estimation. A distance map represents the signed distance of each image voxel to the closest boundary of the segmented region, where 0 values indicate the segmentation boundaries, negative values are for the voxels outside the region and positive values are for the voxels inside.

*2.1.3 Initial model registration*

Before running the model-based level set segmentation, an affine registration is performed to register and match a lung shape model with the input image. The shape model is represented by a mean image and a set of images describing the Principal Components of Variation. It was generated by performing Principal Component Analysis (PCA) on a dataset of 14 lung distance maps, which were derived from lung segmentations from the VISCERAL challenge dataset [22]. The lung shape model's mean image is registered against the obtained lung segmentation distance map (see Section 2.1.2) and the affine transform is saved and used in the following steps of the model-based segmentation.

*2.1.4 Right and left lung segmentation*

The lung segmentation is performed in two sequential steps for each of the two lungs. The segmentation method used in this step is a model-based level set segmentation, presented in [23]. The method combines the threshold-based level set segmentation, described in Section 2.1.2, with a lung model fitting step. This latter step attempts to find the best match of the shape model into the voxel region defined by a previously-set lung look-up table, while also considering a level set curvature constraint. The relative influence in the segmentation between the model and the curvature factor is balanced by using a set of curvature and model weights, which were respectively set to 0.3 and 0.1 in the present work. Moreover, before fitting the model, the previously-computed affine transform (see Section 2.1.3) is used to obtain a rough and preliminary alignment between the shape model and the input CT image to be segmented.

Once the model-based level set segmentation is performed, its two outputs (i.e. one for the left and one for the right lung) are saved as mesh files that can be used as inputs to the subsequent manual editing section of the pipeline.

## 2.2 Manual editing of the lung masks

After automatically segmenting the lungs, a set of manual and interactive segmentation editing tools are available for visual inspection, manual editing and refinement of the lung masks. The two lung regions are loaded as meshes that can be subsequently edited using five available tools, which are presented in the sections below. Two of them operate directly on the mesh points, while the other three work on the corresponding lung mask image. When working on the mask image, the lung meshes are first converted into a lung mask that, after the editing, is converted back into a mesh.

*2.2.1 Magnet tool*

This tool operates directly on the mesh points by displacing them according to the user's mouse clicks. A smooth transition of the displacement of the neighboring points is achieved by modulating the translation of points with a Gaussian function that takes into account the distance between the mouse click and each mesh point: the farther the point is, the smaller its displacement will be. The user can adjust the sigma parameter of the Gaussian function to change the size of the neighborhood that will be affected at each click.

*2.2.2 TPS polyline tool*

This tool operates in a similar way as the magnet tool mentioned above. Only it allows the users to simultaneously use multiple points to deform the mesh. The user can draw one or several polylines defining the 3D profiles where the mesh boundaries should be attracted onto, and the algorithm automatically interpolates and performs the best matching mesh points displacement. These polylines can be drawn in different slices and different views. They are used to create a 3D deformation field using the Thin Plate Spline (TPS) algorithm [24].

*2.2.3 Poly-spline interpolation tool*

When using this tool, the user can create a closed region from scratch by drawing several 2D splines in 3D space. The splines are then automatically interpolated to define a 3D closed surface using a radial basis function approach [25] that is then merged with the existing regions defined by the lung meshes.

*2.2.4 Brush tool*

The brush tool functions as an ordinary painting brush, except it works in 3D. Every mouse click will generate a 3D sphere of a size specified by the bush size. Unlike most other software, the 3D brush in our implementation operates on a distance map instead of binary mask, to ensure smoothness of the resulting mesh after editing. The user can then directly click and drag to paint over the lung mask in order to add or remove lung regions.

*2.2.5 Click-and-drag smart painting tool*

The click-and-drag tool is based on a real-time interactive level set segmentation method, which learns a discrimination function of the object region by analyzing image features around the line segment drawn by the user and the extended line segment beyond the mouse cursor. From the user's perspective, this tool may feel similar to some region-growing tool built in commercial image analysis software. However, since our solution is based on a level-set function, it guarantees local smoothness at the same time. More explanation of this smart painting tool can be found in [26].

## 2.3 COVID-19 lesion segmentation

After manually editing the lung mask, the input resampled image is masked, setting the background value to -2000 HU. Then, another threshold-based level set segmentation is performed to segment the COVID-19 lesions from the masked lungs. However, unlike what was described in Section 2.1.2, a multi-resolution approach was chosen to speed up the global computation time required by this segmentation step.

First, a lower resolution image (down-sampled with an isotropic rescaling of 0.5) is segmented. Then, the segmentation is run again in the high-resolution image by using the low-resolution output as seed region. The threshold range used to identify the COVID-19 lesion regions within the image voxels is -700 to 200 HU. The curvature factor is set to 0.6. The user can also adjust these parameters for each case to achieve a segmentation result that is more visually satisfactory.

Similar to what was performed for the two lung regions, the output lesion segmentations can also undergo a final interactive editing step by following the exact same procedure described in Section 2.2.

## 2.4 Method evaluation

The pipeline presented in this paper was tested on the 10 COVID-19 cases of the open access *COVID-19 CT Lung and Infection Segmentation Dataset* [27]. We excluded the 10 cases from Radiopaedia because their characteristics differ significantly from regular CT images; they have an 8-bit intensity range (instead of X-bit), 3D reconstructed image from these volumes turn out distorted.

Six different annotators were recruited to test the pipeline. They were split into two groups according to their different levels of expertise: a group of three radiologists, which we will refer to as the *Expert Group*, and a group of three annotators with a technical background which we will refer to as the *Novice Group*. All six annotators manually segmented the 10 CT exams using the present tool. Prior to the annotation, all annotators were trained by watching a 15 minutes-long video demo on two COVID-19 cases that are outside the 10 testing cases. The annotators were instructed to include ground-glass opacities, consolidations, reticulation and crazy paving sign in the segmentation, and to include the diffuse part surrounding the lesion.

*2.4.1 Over-all volumetric analysis*

The obtained lesion segmentations were first analyzed by computing the overall lesion volumes for each of the 10 images. These results were compared across the different annotator groups as well as with the available segmentation mask provided by Ma et al. as part of the downloaded dataset [27, 28], which we will refer to as *reference segmentation*. Such volumetric measures allow to preliminary investigate the discrepancy in the results both between and within groups. The discrepancy between groups is indeed proportional to the difference in their mean lesion volumes, while the one within groups is proportional to the standard deviation within the same group.

*2.4.2 Comparison between the Expert and the Novice Group*

We then further compared the Expert and the Novice Groups by following both a volume-wise and a voxel-wise strategy.

As part of the volumetric analysis, we computed the intra-class correlation coefficient (using the ICC(A,1) model) [29], which evaluates the agreement in the volumetric measures of the lesions. This index was calculated for (1) the six annotators together, (2) the Expert Group only and (3) the Novice Group only. Moreover, both the mean absolute and the mean relative lesion volume differences were computed between and within groups. Finally, the linear correlation between the mean volumes and their standard deviations was analyzed.

On the other hand, to perform a voxel-wise comparison, we generated a Novice Group's and an Expert Group's consensus segmentation. These consensus segmentations were derived by performing voxel-wise majority voting with a vote of at least two annotators out of three for each of the two annotator groups. Once the consensus segmentations were obtained, we computed the Dice coefficient [30], the 95$^{th}$ percentile Hausdorff distance [31] and the Jaccard similarity coefficient [32] between them. A further analysis of the spatial overlap between the lesions was then performed by calculating the generalized conformity index (GCI) [29]. Similarly to what was done for the ICC, the GCI was computed both globally and within each annotator group.

*2.4.3 Comparison with the reference segmentation*

The results from both the Expert and the Novice Group were compared also with the available reference segmentations. A Bland-Altman plot was first used to investigate the general volumetric agreement between the reference and the present annotators' segmentations. Moreover, their voxel-wise agreement was analyzed by computing Dice coefficient, Jaccard coefficient and 95$^{th}$ percentile Hausdorff distance. The means and standard deviations of these evaluation metrics were calculated across the segmentation results from all annotators together, as well as separately for each of the two expertise groups.

*2.4.4 Required annotation time*

As a final metric for evaluating the usefulness and validity of the proposed tool, all annotators were asked to provide the time spent using the annotation tool for each of the ten cases. We then analyzed the mean and standard deviation of the time spent to annotate the images, as well as whether it changes according to the level of expertise.

## 3. Results

### 3.1 Volumetric analysis of the segmented lesions

For each of the ten segmented images, we computed the lesion volumes according to all six annotators, as well as to the already provided reference segmentation. The means and standard deviations of such measurements were computed also within each of the two annotator groups, i.e. Expert and Novice (see Table S1 in the Supplementary Material). The plot in Figure 2 represents the volume distribution across groups for every case.

From Figure 2, it can be observed that the discrepancy in the volume measurements between and within groups varies across the different cases. It can be noticed that case 3 and 10 result in a relatively high disagreement both between and within groups. Case 2 and 4, instead, show a particularly higher disagreement within the Expert Group compared to the Novice Group. On the other hand, a higher disagreement within the Novice Group only can be observed for cases 1 and 8.

Figure 3 shows the comparison of the consensus segmentations in some example slices from cases 2 and 10, which resulted, respectively, in a high and a low discrepancy between the results from the Expert and Novice Groups. In case 10, a general tendency of the Novice Group to underestimate the lesion segmentations can be observed.

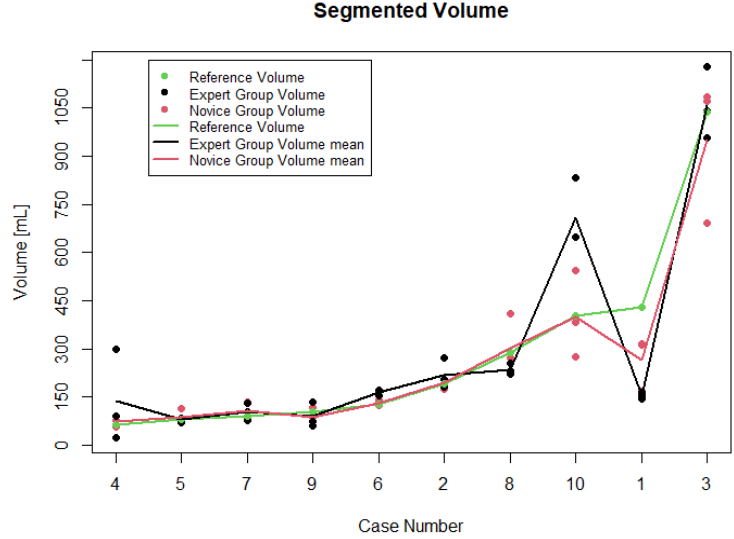

**Figure 2.** For each of the ten segmented cases (*x* axis), their lesion volume (*y* axis) is presented according to each of the annotators (colored dots). The solid lines connect the mean volumes computed within each of the three analyzed group, i.e. the Expert Group (in black), the Novice Group (in red) and the reference segmentation alone (in green). The 10 segmented cases are presented from the lowest (left of the *x* axis) to the highest (right) lesion volume according to the reference segmentation.

### 3.2 Comparison between the two rater groups

The agreement within and between the two annotator groups of the present study was further investigated in order to identify the importance and influence of the radiologists' knowledge to obtain reliable segmentation results. The comparison between the two groups was performed in both a volume-wise and voxel-wise fashion.

*3.2.1 Volume comparison*

The volumetric agreement within and between the two annotator groups of the present study was investigated by computing the intra- and inter-group absolute volume differences, relative volume difference related to the group mean and the ICC coefficient. Table 1 shows that the highest agreement was obtained within the Expert Group (i.e. 0.955), and the lowest within the Novice Group (0.886).

In Figure 4, the relative standard deviations of each group's volumetric measures are plotted against their respective means, showing a correlation between the two. In particular, the two measurements resulted in a linear correlation coefficient of 0.96 for the Novice Group. On the other hand, within the Expert Group, one outlier can be observed (corresponding to case 4), where the standard deviation is higher than the mean (i.e. a relative standard deviation greater than 1 in Figure 4). Such outlier affects the linear correlation coefficient, which results to be equal to 0.58.

*3.2.2 Voxel-wise comparison*

The voxel-wise analysis of the agreement between the two annotator groups was first carried out by using the two consensus segmentations obtained through majority-voting. In particular, such segmentations showed a Dice coefficient of 0.845 ± 0.113 (expressed as mean ± standard deviation), a $95^{th}$ percentile Hausdorff distance of 44.714 ± 21.5 mm, and a Jaccard coefficient of 0.746 ± 0.164.

Furthermore, the GCI was computed to further analyze the degree of spatial overlap between lesions within each of the investigated annotator groups (see Figure 5). A GCI of 0.588 ± 0.155 when performing a global analysis, i.e. considering all six annotators together. On the other hand, the Expert group and Novice Group alone showed a GCI of 0.650 ± 0.195 and 0.545 ± 0.149, respectively. As represented in Figure 6, the resulting GCIs within the two groups do not show a relevant correlation with the lesion volumes.

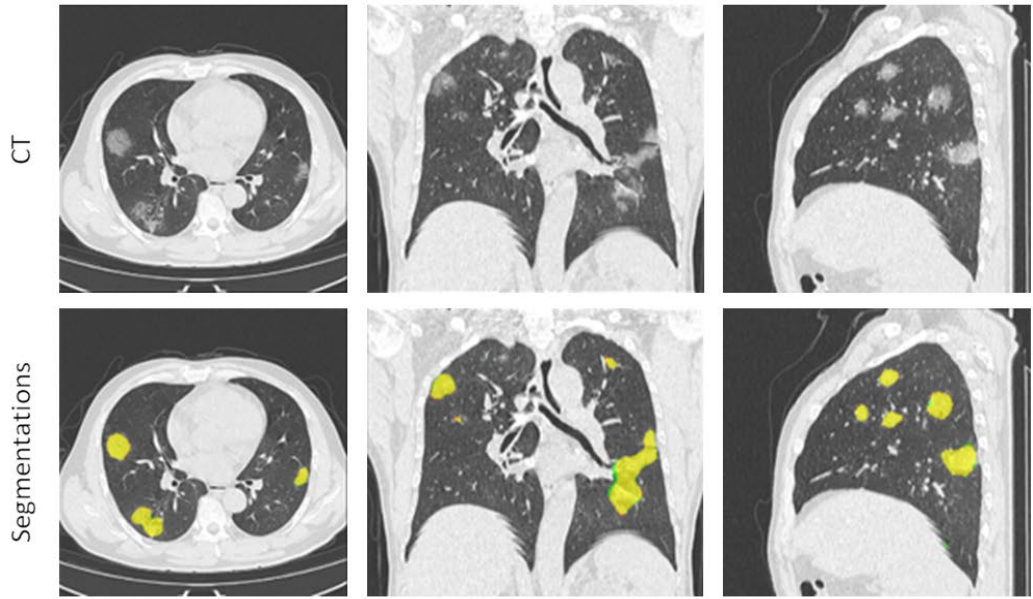

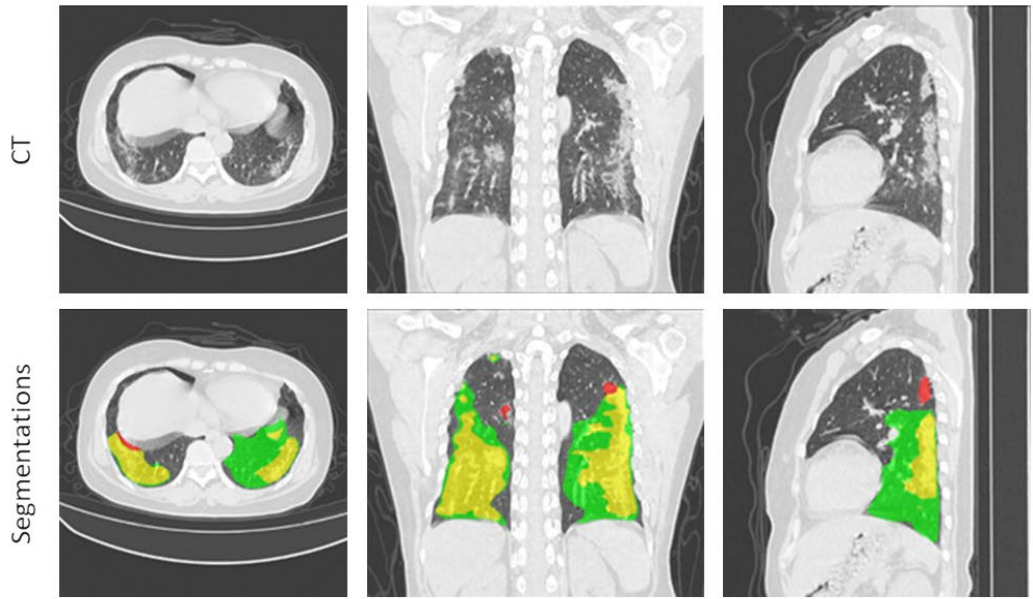

**Figure 3.** Segmentation results in some example slices from all three image views, i.e. axial (left column), coronal (center column) and sagittal (right column). The results are shown for cases 2 (top) and 10 (bottom), which resulted, respectively, in a high and a low agreement between Expert and Novice Groups. For each case, both the original CT image slice and its overlaid segmentation masks are reported. Green areas correspond to regions labeled as lesions only in the Expert consensus segmentation, red areas are labeled as lesions only by the Novice Group consensus, while yellow regions correspond to an intersection between the two groups.

Table 1. Summarized volume measurements across all annotators ("Global" column), across the annotators with a medical background ("Expert Group"), across the annotators with a technical background ("Novice Group"), as well as between the Expert and Novice groups ("Inter-group"). The intra-class correlation coefficient (ICC) and its confidence interval were calculated using the ICC(A,1) model.

| Metric | Global | Expert Group | Novice Group | Inter-group |
| --- | --- | --- | --- | --- |
| **Mean Volume ± standard deviation [mL]** | 278.0 ± 292.2 | 295.6 ± 318.8 | 260.4 ± 267.4 | -- |
| **Mean Absolute Volume Difference [mL]** | 54.7 | 38.4 | 48.9 | 73.0 |
| **Mean Relative Volume Difference** | 23.0% | 18.4% | 17.7% | 24.5% |
| **ICC (A,1) (95% confidence interval)** | 0.899 (0.785~0.969) | 0.955 (0.878~0.987) | 0.886 (0.707~0.967) | 0.926 (0.746~0.981) |

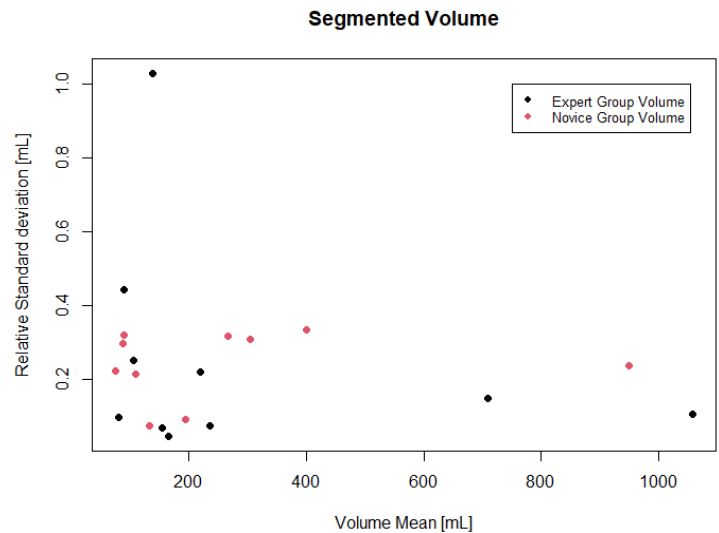

**Figure 4.** The relative standard deviation ($y$ axis) of each of the 10 volumetric measurements within every annotator group (Expert in black and Novice in red) is plotted against its respective mean ($x$ axis) within the same group. The relative standard deviation value was computed by dividing the standard deviation by the respective mean.

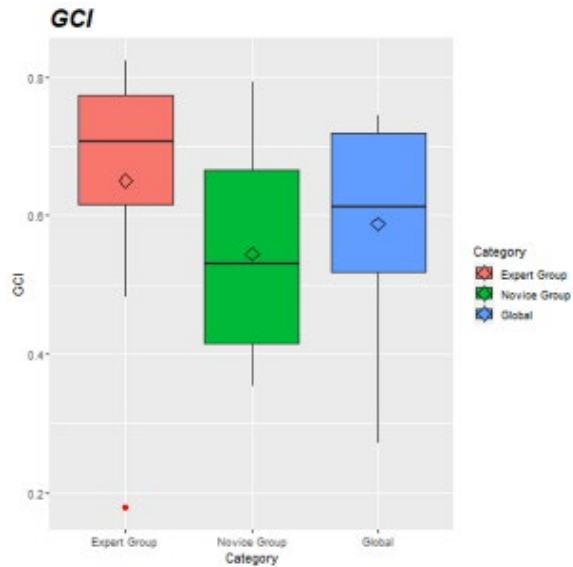

**Figure 5.** Box plots depicting the Generalized Conformity Index (GCI) obtained within only the Expert Group (in red), within only the Novice Group (in green) and within all annotators (in blue).

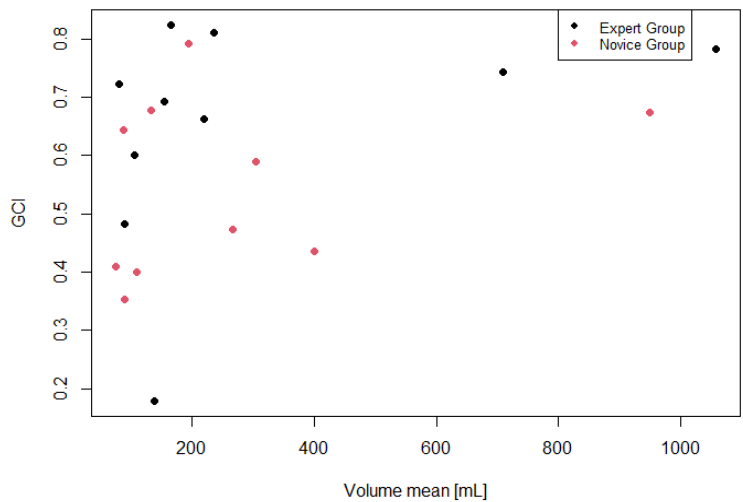

**Figure 6.** The GCI (*y* axis) within each of the two annotator groups (Expert in black and Novice in red) is plotted against the respective mean lesion volume (*x* axis) within the same group.

## 3.3 Comparison with the reference segmentation

The segmentation results obtained from the annotators of the present study were finally compared with the reference lesion segmentation mask that was already provided as part of the dataset.

Figure 7 shows a rather high agreement between the reference segmentation volumes and both the Expert Group and the Novice Group segmentation volumes. Such agreement appears to be stronger when dealing with smaller lesions. Moreover, the biggest volume differences can be observed when comparing the reference with the Expert Group, rather than the Novice Group.

Furthermore, a voxel-wise comparison was also conducted by computing the Dice coefficient, Jaccard coefficient and 95[th] percentile Hausdorff distance between the reference segmentation and the annotator groups analyzed in the present study (see Table 2). All three evaluation metrics showed a slightly better result in the Novice Group compared to the Expert Group, when the reference segmentation is assumed to be the "ground-truth".

### 3.4 Required annotation time

Each annotator recorded the total time spent for segmenting each case, including both the time required to run the semi-automatic pipeline and the manual editing of the annotations using the available manual tools. The overall required annotation time resulted to be equal to 23 ± 12 minutes (expressed as mean ± standard deviation across all raters and all 10 cases). Within the Expert Group, the required time was 23 ± 10 min. On the other hand, the Novice Group reported an annotation time of 22 ± 14 min.

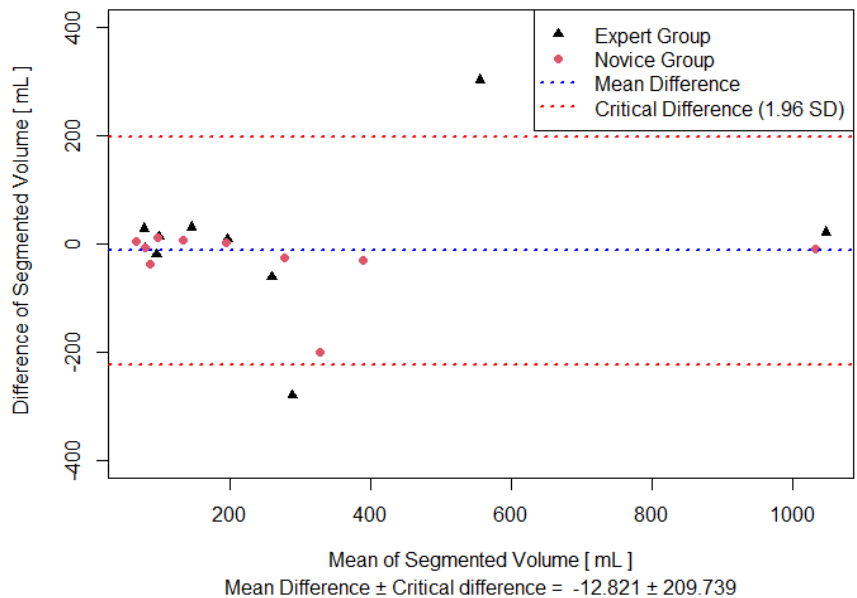

**Figure 7.** Bland-Altman plot describing the agreement between the available reference segmentation and both the Expert Group (in black) and Novice Group (in red) segmentations.

**Table 2.** Voxel-wise agreement between the reference segmentation and (1) all six annotators ("global" column); (2) only the three annotators with a medical background ("expert" column); (3) only the three annotators with a technical background ("novice" column). Dice coefficient, Jaccard coefficient and 95$^{th}$ percentile Hausdorff distance, all expressed as mean ± standard deviation.

| Metric | Global | Expert Group | Novice Group |
|---|---|---|---|
| **Dice coefficient** | 0.704 ± 0.135 | 0.689 ± 0.141 | 0.719 ± 0.129 |
| **Jaccard coefficient** | 0.559 ± 0.150 | 0.542 ± 0.153 | 0.575 ± 0.147 |
| **Hausdorff distance (95$^{th}$ percentile) [mm]** | 59.6 ± 29.1 | 68.8 ± 30.0 | 50.5 ± 25.6 |

# 4. Discussion

In this work, we propose a user-friendly and semi-automatic pipeline for COVID-19 lesion segmentation. This tool showed promising results when used by both expert radiologists and users with a technical background, suggesting the possibility of applying it in a vast range of applications within the scientific community.

Overall, we see a relatively high inter-observer variation among our annotators in our COVID-19 lesion annotation experiments, which is not surprising as COVID-19 lesions tend to have ground-glass appearance with blurry boundaries. This is reflected in both the volume comparison and voxel-wise comparison. Through a questionnaire investigation, we found that most of the annotators are rather confident (average confidence level of 7.13 out of 10) with their annotations. The relatively large disagreement seems to mostly stem from their different perception and interpretation of what a lesion is and where its boundary is defined. Spending more time on each case or using a different tool for the annotation will not likely reduce the inter-observer variation greatly. However, spending more time training the annotator to achieve better consensus on their interpretation of the lesion boundary might help to reduce the differences. On the other hand, our experiment setup with less adequate training will be closer to the clinical reality. Our reported inter-observer variation on COVID-19 lesion segmentation could provide reference for future studies that aim to build large scale annotated COVID-19 chest CT databases. Also, future AI researchers should be aware of the rather high inter-observer variation associated with their training data when using these annotated images to train AI models. Moreover, clinicians should be aware of these uncertainties when using the quantitative COVID-19 lesion burden estimation produced by automated or semi-automated software to make their clinical evaluation of the patient's status.

## 4.1 Comparison between the Expert and the Novice Group

One focus of our quantitative evaluation of the segmentation results is to analyze the agreement in the segmentation results between and within two groups of annotators: the Expert Group, made up of three expert radiologists, and the Novice Group, consisting of three engineers lacking of a solid radiology knowledge.

By performing a volumetric analysis of the segmented lesions, we could come to two major observations. The first one is that there is no visible and consistent trend in the segmentation results of the two groups, since the discrepancy in volume measurements varies largely across the ten tested cases. This can be particularly observed from the measurements presented in Figure 2 and Table S1, as well as from the Bland-Altman plot in Figure 7. It is indeed noticeable that none of the two groups obtains consistently larger or smaller volumes, also when compared to the reference segmentation in the latter figure.

The second observation that we could see from our volumetric analysis is that satisfactory agreement could be obtained both between and within the groups. As expected, the Expert Group showed the highest within-group agreement (with an ICC of 0.955), reflecting their common expertise and background knowledge. However, the Novice Group also showed a rather high ICC (i.e. 0.886), which suggests that the proposed tool can help even non-expert users to achieve a good level of agreement. This hypothesis is further confirmed by the high ICC observed both globally and between the two groups, i.e. the present pipeline could allow users with different levels of expertise to achieve segmentation results that are in rather good agreement with those delineated by a radiologist. On the other hand, one factor seemed to consistently affect the within group-agreements: lesions with bigger volumes were associated with a higher standard deviation within each group. This can be noted from Figure 4, which shows that the relative standard deviation (i.e. normalized by the mean lesion volume) is relatively constant with respect to the lesion volume. However, this affected both novices and experts, showing that such difficulty is hard to overcome even when solid radiologic knowledge is present.

The analysis of the agreement between the two groups was then also performed in a voxel-wise fashion by using the two consensus segmentations. We are aware of the challenges of both performing accurate lesion segmentations (especially along the lesion boundaries) and of relying on just one consensus segmentation to represent a whole expertise group. However, an average Dice coefficient of 0.845 was achieved, which we believe to indicate a very promising result and a high degree of spatial overlap between the two groups across the analyzed cases.

The voxel-wise agreement was investigated also within groups by measuring their GCIs. Similarly to what was observed for the volumetric analysis, the Expert Group showed a higher agreement than the Novice Group also in terms of spatial overlap. This was again expected due to the level of expertise of the annotators. However, in contrast to the volumetric agreement, we could notice that the level of spatial agreement is much less correlated to the lesion volume (see Figure 6). Thus, we think that the main challenges of delineating geometrically accurate lesions are more related to factors such as image quality, disease stage and the rater's personal perception than to the simple lesion volume.

## 4.2 Comparison with the provided reference segmentations

The dataset used in the present study included, together with the original CT scans, also their respective lung and lesion segmentations performed by a team of radiologists [28]. Therefore, we decided to compare such reference segmentations with those produced by the annotators enrolled for the present study.

Comparing the lesion volume measurements from our annotators with those from the reference segmentation, we could observe that their agreement is higher when dealing with smaller volumes. This is consistent with what was observed in Section 4.1 regarding the comparison between Expert and Novice Group in terms of lesion volumes. Thus, this further strengthens the argument that the more spread the lesions are, the less reliable their volume estimations may become.

We then computed three voxel-based evaluation metrics (Dice coefficient, Jaccard coefficient and Hausdorff distance), using the manual reference segmentation as "ground truth". All three metrics—both when computed globally on all six annotators and when computed group-wise—indicated less spatial overlap compared to the analysis between our Expert Group and Novice Group using our proposed tool. This suggests that this tool guides the annotators—no matter what their expertise is—to achieve results that are particularly consistent with each other. We indeed found particularly relevant and interesting that our enrolled radiologists produced segmentations that appear to be more in agreement with those from the Novice Group, compared to those manually produced by another group of radiologists. It is therefore important to be aware that the segmentations produced in the reference study [28] could also be influenced by the radiologists' perception and training, making it hard to point out which segmentations are the most accurate.

We believe that the above-discussed observations show how hard it is to obtain reliable and consistent lesion segmentations, even when performed by expert radiologists. This indicates how the analysis of inter-rater agreement is fundamental for these applications and that a certain bias in the segmentation results cannot be easily eliminated. However, even if results can be affected by erroneous estimations and biases, our tool was shown to succeed at least in trying to reduce the discrepancy between different levels of expertise by providing a guided and user-friendly workflow.

## 4.3 Applicability of the proposed tool

Given the promising results observed in this study, we believe that the present tool can be useful for analyzing chest CT images of COVID-19 patients in several ways.

First of all, the possibility of segmenting both lungs and lesions together allows for measuring also the well-aerated lung volume, which can predict adverse outcome or death in COVID-19 patients [33]. Thus, our pipeline might allow risk stratification of hospitalized patients.

Another application is the generation of annotated training data, aiming to implement new machine learning-based and deep learning-based tools. Our tool can indeed aid the further automation of new segmentation approaches by providing lesion masks that can be used as "ground-truth" during the training process. Moreover, it can also allow the training of new automated diagnostic tools by directing them towards regions presenting important image features.

Apart from COVID-19-centered applications, we also believe that the proposed tool could be adapted for analyzing other lower respiratory tract infections or other diseases affecting the lung parenchyma, such as chronic obstructive pulmonary disease, interstitial lung disease, or tuberculosis.

All above-discussed applications are facilitated by the fact that the pipeline is partially automatic and partially involves manual adjustments. The possibility of manually adjusting the segmentation results throughout the workflow allows to obtain more reliable and controlled lung and lesion masks. On the other hand, the automated steps of the pipeline have two other main advantages. The first one is to provide less-expert users with more guidance on what may actually constitute a lesion and what not. By receiving a preliminary automatic segmentation, it is indeed going to be less likely for the user to miss certain lesions or to be unsure about the lesion boundaries. The second advantage of including automated steps into the pipeline is speed, since the annotators reported an average annotation time of 23 minutes. This average value was also equal to 23 minutes for the enrolled radiologists only, who represent the expertise group that would normally be involved in the manual delineation of lesion masks. In contrast, the time spent for manually generating the reference segmentations—which were included as part of the dataset employed in the present study—had been reported to be equal to $400 \pm 45$ minutes per scan [28]. Thus, this suggests that the proposed tool can significantly improve annotation efficiency.

### 4.4 Future work

In the future, we aim to further improve the functionalities and efficiency of the proposed pipeline in order to optimize the segmentation process further.

The present version of the segmentation tool uses pre-set values for the thresholds and smoothness parameters used in the automatic segmentation steps. Such fixed values were set purely empirically and are always the same for each input image. However, in the future we would like to make the selection of such values automatic and based on the statistical distribution of the voxel intensities. This would help to reduce the amount of manual tweaking required from the user.

Another possible future extension may consist in including an automatic registration of a reference lung shape into the lung of the subject being segmented. In this way, lesion locations could be related to specific lung segments. This would allow to compare the lesion distributions between different patients and to facilitate studying disease progression over time.

## 5. Conclusions

In this work, we presented an interactive and user-friendly software for segmenting COVID-19 lesions from input chest CT scans, implemented within the software Mialab (http://mialab.org).

The use of our proposed tool, which alternates automatic segmentation steps with manual corrections, led to promising results when employed by both radiologists and users with an engineering background. The proposed pipeline was shown to aid and speed up the process of creating lesion segmentation masks.

We also evaluated the inter-observer segmentation variability and how it can be influenced by several factors, ranging from the lesion volume to the rater's personal interpretation of where the lesion boundaries should be drawn. Our analysis could provide guidance and reference for future efforts to build large COVID-19 chest CT databases for AI model training. It indeed highlights that both AI researchers and clinical doctor should be aware of the uncertainty associated with the annotated image data and the quantitative volumetric measurements.

# Acknowledgements

This work was supported by the Swedish Heart-lung-foundation (grant 2019-0540), the Swedish Research Council (grant 2018-04375), the European Union's Horizon 2020 research and innovation program under the Marie Sklodowska-Curie grant agreement No 795747 and donations from Einar Mattsson AB and the Allba Foundation. M.P. discloses financial interests in Prismatic Sensors AB and research collaboration with GE Healthcare.

# Supplementary material

**Table S1.** For each of the ten segmented CT images (table rows), the respective lesion segmentation volumes (in mL) are reported. Such volumes are expressed as mean ± standard deviation and computed across: (1) all six annotators (second column); (2) only the three annotators with a medical background (third column); (3) only the three annotators with a technical background (fourth column). In the fifth column, the lesion volumes from the available reference segmentations are also reported for comparison.

| Case | Global | Expert Group | Novice Group | Reference |
|---|---|---|---|---|
| 1 | 210.119 ± 80.725 | 154.926 ± 10.692 | 265.312 ± 83.892 | 428.545 |
| 2 | 206.762 ± 35.24 | 219.147 ± 48.32 | 194.376 ± 17.6 | 191.945 |
| 3 | 1003.6 ± 169.21 | 1058.475 ± 111.788 | 948.725 ± 223.717 | 1036.595 |
| 4 | 106.477 ± 97.067 | 138.306 ± 142.265 | 74.649 ± 16.635 | 63.729 |
| 5 | 83.19 ± 17.399 | 79.782 ± 7.724 | 86.597 ± 25.736 | 82.407 |
| 6 | 148.92 ± 19.423 | 165.163 ± 7.558 | 132.678 ± 9.724 | 129.535 |
| 7 | 106.757 ± 22.31 | 104.758 ± 26.337 | 108.755 ± 23.211 | 91.759 |
| 8 | 270.012 ± 70.658 | 236.014 ± 17.336 | 304.011 ± 93.343 | 290.027 |
| 9 | 89.126 ± 30.919 | 89.726 ± 39.782 | 88.527 ± 28.394 | 104.767 |
| 10 | 555.178 ± 200.652 | 709.605 ± 105.772 | 400.751 ± 133.875 | 403.777 |